\documentclass[conference]{IEEEtran}
\usepackage{caption}

\usepackage{mdwlist}
\usepackage{cite}
\usepackage{amsfonts}
\usepackage{amssymb}
\usepackage[cmex10, fleqn]{amsmath}
\usepackage{color}
\usepackage{algorithm}
\usepackage{algorithmic}
\usepackage{multirow}
\usepackage{bm}
\usepackage{booktabs}
\usepackage{graphics}
\usepackage{graphicx}
\usepackage{ifpdf}
\ifpdf
  \usepackage{epstopdf}
\fi

\usepackage{float}
\usepackage{epsfig}
\usepackage{multirow}
\usepackage{enumerate}

\usepackage{nomencl}
\makenomenclature

\usepackage{arydshln}

\usepackage{verbatim}
\usepackage{tikz,times}
\usetikzlibrary{calc, shapes, backgrounds}

\makeatletter
\newcommand\figcaption{\def\@captype{figure}\caption}
\makeatother

\ifCLASSINFOpdf
\else
\fi
\IEEEoverridecommandlockouts
\begin{document}

%
\title{ Hybrid Optimization Method for Reconfiguration of AC/DC Microgrids in All-Electric Ships }
%

\author{\IEEEauthorblockN{~Qimin~Xu,~Bo~Yang,~Zhizhang~Pan,~Feilong~Lin,~Qiaoni~Han,~Cailian~Chen,~Xinping~Guan }
\IEEEauthorblockA{Department of Automation, Shanghai Jiao Tong University, Shanghai, China\\
Collaborative Innovation Center for Advanced Ship and Deep-Sea Exploration, Shanghai, China\\
Key Laboratory of System Control and Information Processing, Ministry of Education of China, Shanghai, China} \\
Email: \{qiminxu, bo.yang, cailianchen, and xpguan\}@sjtu.edu.cn
\thanks{This work was supported by National Key Research and Development Program of China (2016YFB0901901), National Natural Science Foundation of China (61221003, 61174127, 61104033, and 61273181) }}%


%


\maketitle

\begin{abstract}
Since the limited power capacity, finite inertia, and dynamic loads make the shipboard power system (SPS) vulnerable, the automatic reconfiguration for failure recovery in SPS is an extremely significant but still challenging problem. It is not only required to operate accurately and optimally, but also to satisfy operating constraints.
In this paper, we consider the reconfiguration optimization for hybrid AC/DC microgrids in all-electric ships.
Firstly, the multi-zone medium voltage DC (MVDC) SPS model is presented. In this model, the DC power flow for reconfiguration and a generalized AC/DC converter are modeled for accurate reconfiguration.
Secondly, since this problem is mixed integer nonlinear programming (MINLP), a hybrid method based on Newton Raphson and Biogeography based Optimization (NRBBO) is designed according to the characteristics of system, loads, and faults. This method facilitates to maximize the weighted load restoration while satisfying operating constraints.
Finally, the simulation results demonstrate this method has advantages in terms of power restoration and convergence speed.
\end{abstract}



%
\IEEEpeerreviewmaketitle

\section{Introduction}

Shipboard power system (SPS) can be considered as an isolated microgrid, because it is self-powered by distributed electrical power generators \cite{hebner2015technical}. 
Comparing with terrestrial systems, the reliability, safety and fault tolerance design for all-electric ships (AES) is more rigorous. There are three salient features as follows.
Firstly, the consequences of a minor fault in a system component can be catastrophic due to the intensive coupling feature of SPS \cite{zhang2008bibliographical};
Secondly, system failure is more fatal for ships than for terrestrial systems, because the personal safety on board is more endangered \cite{hebner2015technical};
Lastly, dynamic loads constitute a large proportion of ship power systems, which changes with operation mode switching.
Therefore, effective and intelligent reconfiguration of SPS is essential in response to electric plant casualties and mission changes of the ships.

From the perspective of electrical design of AES, there are three architectures of SPS to date, i.e., medium voltage DC (MVDC), AC (MVAC), and higher frequency AC (HFAC). As the ever-increasing DC-based loads, it is likely that AES will feature a medium voltage primary distribution system in the future \cite{hebner2015technical}. 
Many works have been done on the optimal SPS reconfiguration in MVDC AES. The objectives include reducing the operating cost, maximizing either the weighted power or the current supplied to loads \cite{Bose2012Analysis,das2013dynamic}, maximizing stability margins \cite{arcidiacono2012generation}, minimizing the number of switch operation \cite{jiang2012novel}, etc. 
All of these works only focused on load management, but was no consideration of the impact of ship operation. 
Additionally, since the MVDC SPS is a converter based electric power system and the converter power losses make up the most of its power loss, converter model is needed for accurate control and system level analysis. For example, as shown in \cite{qi2014integrated}, the maximum converter and transformer power losses in MVDC and MVAC can be up to 1102 kW and 1474 kW, respectively. Therefore, if the power loss cannot be well considered, a invalid control decision would be made, which would result in electrical energy quality descend or even system instability.

On the other hand, for the optimal reconfiguration of SPS without speed constraint and converter model, various algorithms have been proposed with different methodologies.
In \cite{amba2009genetic,kumar2007shipboard,Padamati2007Application,jiang2012novel}, evolutionary algorithms including genetic algorithm (GA) and particle swarm optimization (PSO) are proposed for the optimal SPS reconfiguration. To the same problem, an interior-point based method was proposed in \cite{Bose2012Analysis}. 
A reinforcement learning based algorithm was proposed in \cite{das2013dynamic}, which also considers the optimal sequence of switching operation.
Most of these works focus on improving the restored power under the operating constraints. 
However, in IEEE Std. 45-2002 \cite{ieee2002shipboard} the maximum frequency deviations of $ \pm $ 3\% for continuous operation and $\pm$ 4\% for transients must be less than 2 seconds. The SPS reconfiguration delay is particularly significant to guarantees the ship's survivability.
Time constrain of algorithms must draw enough attention, especially when growing loads increases the complexity of problem.

Taken these considerations into account, firstly we formulate the comprehensive model of MVDC SPS including converter model and multi-zone DC power flow model.
The formulated optimal reconfiguration problem is maximizing the weight load power considering generator, load, AC and DC power flow constraints, which is an mixed-integer non-linear programming (MINLP) with integer variables (load switches and redundancy switches) and continous variables (active and reactive power of generators).
Secondly, to solve this problem, we design a hybrid algorithm based on Newton Raphson and Biogeography based Optimization (NRBBO) method to realize fast and effective reconfiguration. Specifically,  (i) \textbf{Decoupling}:
The DC and AC power flows are decoupled and calculated iteratively, thus the orignal problem is converted to mixed-integer linear programming (MILP) problem and non-linear programming (NLP) problem. BBO is responsible for DC part, while NR is for AC and converter part. 
(ii) \textbf{Mode distinction}: Three fault modes are defined by the fault position. Based on DC power flow formulation, the optimal solutions of three modes can be obtained respectively with reduced complexity.
(iii) \textbf{Layer search}: The switch variables used for load control are layered according to load priority, thus each search set consisting of switch variables are reduced greatly.
The mode processing and layered methods used for reconfiguration does not impose any restrictions on the topology of SPS.
Thirdly, the relationship between the restored power and the position and number of faults is analyzed, which aims at quantifying the system restoring ability against faults and giving a suggestion for structure design.

The paper is organized as follows: in Section \ref{sec:model}, MVDC SPS is modeled and the optimization problem is formulated; 
Section \ref{sec:reconfiguration} details the proposed NRBBO method; the performance of our method is evaluated and compared with existing reconfiguration methodologies in Section \ref{sec:simulation}. Finally, the conclusion is drawn in Section \ref{sec:conclusion}.

\section{Model of Shipboard Power System}
\label{sec:model}
\begin{figure}[ht]
\setlength{\belowcaptionskip}{-0.8cm}
  \centering
    \includegraphics[width= 0.5 \textwidth]{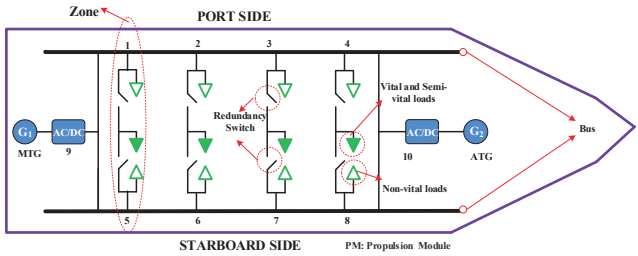}
  \caption{MVDC shipboard power system architecture.}
  \label{fig:arch}
\end{figure}

In this section, the system models of SPS are described in detail. 
The classic architecture of MVDC SPS with $ K $ electric zones is demonstrated in Fig.\ref{fig:arch}. The DC zones are powered by a starboard bus (SB) and a port bus (PB) which connect to $ M $ converters and generators. There are two type generators: main generator (MG) and auxiliary generator (AG). 
The main notations used in this work are summarized in Table \ref{tab:1}.
\begin{table}
  \centering
  \caption{Main notations}
  \label{tab:1}
  \footnotesize
  \begin{tabular}{p{2cm}|p{6cm}}
  \hline
  Notation    & Physical interpretation \\
  \hline
  \hline
  $ \mathcal{L} $, $ L $, $ l $ & Set, number and index of loads \\
  $ \mathcal{K} $, $ K $, $ k $ & Set, number and index of DC zones \\
  $ M $; $ m $, $ n $ & Number and indexes of generators and converters \\
  $ N $; $ i $, $ j $ & Number and indexes of DC buses \\
  $ \mathcal{H}_h $ & Set of the $ h $-th habitat(solution), $ h \in \{1,2,\cdots,H\} $  \\
  $ Z $, $ z $ & Number, index of species(switches), $ Z=L $ \\
  $ G $, $ g $ & Number, index of layers \\
  $ G_s $ & Layer number that the BBO starts from \\ 
  $ F_{{po}_{i,j}} $ & Fault position between $ i $-th and $ j $-th DC buses \\
  $ \mathcal{F}_{po} $ & Set of fault positions  \\  
  $ \mathrm{max} $, $ \mathrm{min} $ & Superscript denoting minimum and maximum \\
  $ \mathcal{B}_i $ & Set of loads at the $ i $-th DC bus, $ \mathcal{B}_i \subset \mathcal{L} $ \\
  $ S_{l} $ & Switch of the $ l $-th load, $ s_{l} \in \{0,1\} $  \\
  $ \mathcal{S} $, $ \bm{S}_i $ & Set of all switches and the switches at the $ i $-th DC bus, $ \bm{S}_i = \{ s_l \colon l \in \mathcal{B}_i \} $  \\
  $ I_{load_l} $, $ \bm{I}_{load_i} $ & Current of the $ l $-th load, and set of load currents in the $ i $-th DC bus, $ \bm{I}_{load_i} = \{ I_{load_l} \colon l \in \mathcal{B}_i \} $  \\  
  $ {I}_{in_i} $, $ {I}_{b_i} $ & Injected current and total load current in the $ i $-th DC bus \\
  $ w_{V_g} $ & Weight factor of the $ g $-th grade loads, and there are  three grades loads: vital, semi-vital, non-vital \\ 
  $ P_{l} $, $ w_{l} $ & Power and weight factor of the $ l $-th load, $ w_{l} \in \{w_{V_1}, w_{V_2}, w_{V_3} \} $ \\
  $ Y_{dc_{ij}} $ & Branch admittance between the $ i $-th and  $ j $-th DC buses \\
  $ \mathrm{Y}_{dc} $ & DC admittance matrix \\
  $ U_{dc_i} $, $ I_{dc_i} $ & Voltage and input current at the $ i $-th DC bus  \\
  $ P_{ac_m} $, $ Q_{ac_m} $ & Active and reactive power of $ m $-th generator \\
  $ U_{ac_m} $, $ \delta_{ac_m} $ & Voltage and angle of the $ m $-th AC bus \\
  $ I_{ac_{mn}} $, $ Y_{mn} $ & Current and branch admittance between the $ m $-th and $ n $-th AC bus \\
  $ \delta_{m} $ &  Angle associated with the voltage at the bus $m$ \\ 
  $ P_{C_m} $, $ Q_{C_m} $ & Active and reactive input power of $ i $-th converter \\
  $ U_{C_m} $, $ I_{C_m} $ & Input voltage and current of $ m $-th converter \\
  $ P_{cpl_m} $, $ P_{loss_m} $ & Constant and total power loss of $ m $-th converter \\
  $ P_{oc_m}  $ & Output power of $ m $-th converter \\
  $ S_{P_k} $, $ S_{S_k} $ & Redundancy switches of PB and SB in the $ k $-th zone, ${S}_{P_k}, {S}_{S_k} \in \{0,1\}$ \\
  $ \mathcal{P}_V, \mathcal{P}_{SV}, \mathcal{P}_{NV} $ & Sets of vital, semi-vital, non-vital loads' power \\
  $ P_{V_g} $ & Power of one load in the $ g $-th grade, $ P_{V_1} \in \mathcal{P}_V, P_{V_2} \in \mathcal{P}_{SV}, P_{V_3} \in \mathcal{P}_{NV} $ \\
  $ P_{C_m}^{(t)} $ & Active Power of converter in the $ t $-th iteration, $ t \in \{1,2,\cdots,T\} $ \\
    \hline
  \end{tabular}
\end{table}

\subsection{AC Power Flow Model}
The power flows in SPS can be modeled using the branch ﬂow model
\begin{align} 
\label{P_flow} & P_{ac_{m}} = U_{ac_m} \sum_{n=1}^{M} U_{ac_{n}} Y_{mn}Re\{\delta_{m}-\delta_{n}\} \\ 
\label{Q_flow} & Q_{ac_{m}} = U_{ac_m} \sum_{n=1}^{M} U_{ac_n} Y_{mn}Im\{\delta_{m}-\delta_{n}\},
\end{align}
where $ P_{ac_m} $, $ Q_{ac_m} $, $U_{ac_m}$ , and $\delta_{m}$ denote the active power, reactive power, voltage, and phase angle at the AC bus $m$. In each bus, $P_{ac_m}= P_{g_m} - P_{d_m}$, where $P_{d_m}$ denotes the load demand at the AC bus $m$. $Y_{ac_{mn}}$ represents the branch admittance between the AC bus $m$ and $n$.

The generator at each AC bus also need to satisfy the following constraints: 
\begin{align}
\label{PG_limit} & P_{g_m}^\mathrm{min} \leq P_{g_m} \leq  P_{g_m}^\mathrm{max} \\ 
\label{PQ_limit} & Q_{g_m}^\mathrm{min} \leq Q_{g_m} \leq  Q_{g_m}^\mathrm{max} \\ 
\label{V_limit} & U_{ac_m}^\mathrm{min} \leq U_{ac_m} \leq  U_{ac_m}^\mathrm{max} \\
\label{I_limit} &  I_{ac_{mn}} \leq  I_{ac_{mn}}^\mathrm{max} \\
\label{delta_limit}& \delta_{m}^\mathrm{min} \leq \delta_{m} \leq  \delta_{m}^\mathrm{max}. 
\end{align}

\subsection{Converter Loss Model}
%
%
The power loss of converter $ m $ has three parts, namely, the constant part $ P_{cpl_m} $, the linear and the quadratical parts. 
The latter two depend on the current $ I_{C_m} $, which can be calculated by the input power $ P_{C_m} $, $ Q_{C_m} $ and voltage $ U_{C_m} $. Thus, the power loss model of converter \cite{beerten2012generalized} can be expressed as
\begin{align}
\label{P_loss}
& P_{loss_m} = P_{cpl_m} + a\cdot I_{C_m} + b\cdot I_{C_m}^{2}, 
\end{align}
\begin{align}
\text{with} \quad  I_{C_m}  = \dfrac{\sqrt{P_{C_m}^{2}+Q_{C_m}^{2}}}{\sqrt{3}U_{C_m}},
\end{align}
where $ P_{cpl_m} $, and coefficients $ a $ and $ b $ are related to the electrical feature of the converter. Moreover the relationship between output power $ P_{oc_m} $ and input power $ I_{C_m} $ can be written as
\begin{align}
\label{eqn:out_power}
 P_{oc_m} = P_{C_m}-P_{loss_m}.
\end{align}

The current and active power of converter model are bounded as follow: 
\begin{align} 
\label{IC_limit}&  I_{C_{m}} \leq  I_{C_{m}}^\mathrm{max} \\ 
\label{PC_limit} & P_{oc_m}^\mathrm{min} \leq P_{oc_m} \leq  P_{oc_m}^\mathrm{max}.
\end{align}

\subsection{DC Zone Power Flow Model}
In the DC zone, there are $ N $ buses in total. 
The power flow can be expressed in a similar way as a general AC power flow. The injected current at the bus $ i $ can be written as the sum of the total load currents in this bus and the total current flowing to other $ N - 1 $ buses. Since loads are controlled by corresponding switches, the current equation at the bus $ i $ is given as:
\begin{align}
\label{current_load_in_bus}
I_{in_i}=\sum_{l \in \mathcal{B}_i}{s_{l} \cdot I_{load_{l}}} + \sum_{\mbox{\tiny$\begin{array}{c}
j=1\\
j\neq i \end{array}$}}^{N} 
{Y}_{dc_{ij}}\cdot (U_{dc_i} - U_{dc_j}),
\end{align}
where $ \mathcal{B}_i $ is the set of loads powered by the bus $ i $. $ I_{load_l} $ and $ s_{l} $ indicate the current and switch status of load $ l $ at the bus $ i $. $ s_{l} $ is a binary variable.

Here, for convenience we set the loads as constant current components for analysis. They can also be set as constant impedance or power components.
\begin{figure}[ht]
\begin{center}
\setlength{\belowcaptionskip}{-0.3cm}
\includegraphics[width= 0.5\textwidth]{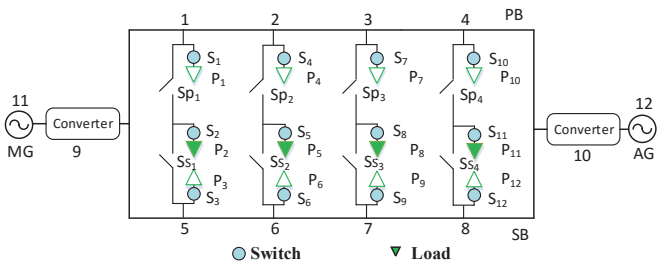} 
\caption{ Ship power system model. } 
\captionsetup{justification=centering} 
\label{fig:normal} 
\end{center} 
\end{figure}

In order to simplify the DC admittance matrix, the DC buses are numbered like Fig.\ref{fig:normal}. The buses in PB and SB are numbered in order respectively, and the buses connected to the converter are numbered in the last.

The redundancy switches in each zone is mutually exclusive, which determine if the vital and semi-vital loads is powered by PB or SB. The constraint is written as: 
\begin{align} 
\label{eqn:redundence_switch} & {S}_{P_k} + {S}_{S_k} =1, \quad {S}_{P_k}, {S}_{S_k} \in \{0,1\}, {k} \in \mathcal{K},
\end{align}
thus ${S}_{P_k}$, ${S}_{S_k}$ determine the load sets $\mathcal{B}_i$ at corresponding bus. 

Combining all the current equations in (\ref{current_load_in_bus}) result in
\begin{align}
\label{switch_power}
\bm{I}_{in} = 
\left[ \begin{array}{c}
\bm{S}_{1} \bm{I}_{load_1} \\
\vdots\\
\bm{S}_{\hat N} \bm{I}_{load_{\hat N}}\\
\hline
0\\
\vdots\\
0
\end{array} \right]
 +\mathrm{Y}_{dc} \bm{U},
\end{align}
where $ \bm{I}_{in} = \left[ I_{in_1} \ I_{in_2} \cdots I_{in_N}  \right] ^\mathrm{T} $ denotes the current vector, $ \bm{S}_{i} \bm{I}_{i} $ the total load current at the bus $ i $, $\hat N = N-M$, 
and $ \bm{U} = \left[ U_{dc_1} \ U_{dc_2} \cdots U_{dc_N}  \right] ^\mathrm{T} $ the voltage vector. 
Assuming a unipolar DC grid, the active power injected in the bus $ i $ from AC grid can be expressed as
\begin{align}
\label{dcpower_cal}
P_{oc_i} = U_{dc_i} I_{dc_i}, \quad \forall i \leq M,
\end{align}

Combining (\ref{switch_power}) and (\ref{dcpower_cal}), the DC power can be written as
\begin{align}
\left[ \begin{array}{c}
0\\
\vdots\\
0\\
\hline
{P_{oc_{\hat N+1}}}/{U_{dc_{\hat N+1}}} \\
\vdots\\
{P_{oc_N}}/{U_{dc_N}} \\
\end{array} \right]
=
\left[ \begin{array}{c}
\bm{S}_{1} \bm{I}_{{load}_{1}} \\
\vdots\\
\bm{S}_{{\hat N}} \bm{I}_{load_{\hat N}}\\
\hline
0\\
\vdots\\
0
\end{array} \right]
+
\mathrm{Y}_{dc} \bm{U},
\end{align}
Under normal conditions, we assume that one converter bus is the DC slack bus. Here converter $N$ is set as slack bus, the undetermined variables contain $ U_{dc_1}, \cdots, U_{dc_{N-1}} $ and $ P_{oc_N} $. The simplified equations is written as
\begin{align}
\label{P_DC_cal} 
\left[
-I_{b_1} \  \cdots -I_{b_{N-M}} \ \frac{P_{oc_{N^{*}}}}{U_{dc_{N^{*}}}} \ \cdots \ \frac {P_{oc_N}}{U_{dc_N}} \right] ^\mathrm{T}
= \mathrm{Y}_{dc}\cdot \bm{U},
\end{align}
where $ I_{b_i} $ denote total load current at the $ i $-th DC bus. The detail equations are written as
\begin{equation}
\label{total_dcpower}
\left[ \begin{array}{c}
-I_{b_1}\\
\vdots\\
-I_{b_{N-M}}\\
P_{oc_{N^{*}}}/U_{dc_{N^{*}}} \\
\vdots\\
P_{oc_{N-1}}/U_{dc_{N-1}} \\
\hline
P_{oc_N}/U_{dc_N} \\
\end{array} \right]
=
\left[ \begin{array}{c|c}
\mathrm{Y}_{dc_{11}} & \bm{Y}_{dc_{12}}\\
\hline
\bm{Y}_{dc_{21}} & y_{dc_{22}}
\end{array} \right]
\left[ \begin{array}{c}
U_{dc_1} \\
\vdots\\
U_{dc_{N-1}}\\
\hline
U_{dc_N}
\end{array} \right].
\end{equation}
where the matrix $ \mathrm{Y}_{dc_{11}} $ is $ (N-1)\times(N-1) $ dimensional matrix, $ \bm{Y}_{dc_{12}} $ and $ \bm{Y}_{dc_{{21}}} $ are $ N - 1 $ dimensional column and row vector, and $ y_{dc_{22}} $ is a scale. It is divided into two parts. Therefore, the DC power flow can be calculated iteratively by (\ref{total_dcpower}). 

Additionally, the current and volatage are meet the constraints as follow:
\begin{align} 
\label{U_cons} & U_{dc_i}^\mathrm{min} \leq U_{dc_i} \leq  U_{dc_i}^\mathrm{max} \\
\label{I_cons} & I_{ij} \leq  I_{ij}^\mathrm{max}. 
\end{align}

\subsection{Load Model}
The loads are powered by a set of buses which run longitudinally along the PB and SB. The circuit breaks (switches), which can be used to connect the loads and buses.
There are three kinds of electric loads: vital loads $ \mathcal{P}_{V}$, semi-vital loads $\mathcal{P}_{SV}$ and non-vital loads $\mathcal{P}_{NV}$. At a certain time vital and semi-vital loads can only be powered by one bus (PB or SB), which is determined by the redundancy switches. This redundency topology aims to improve the stability of power supply for vital and semi-vital loads. Non-vital loads only connect to one bus, PB or SB. 
For the safety operation of vital loads, it is necessary to guarantee the power supply of vital loads all the time, which is described as follow:
\begin{align} 
\label{eqn:vital_load} 
\sum_{m \in \mathcal{M}} P_{oc, m} - \sum_{l \in \mathcal{P}_{V}} P_{l} \geqslant 0.
\end{align}

  
\subsection{Reconfiguration Optimization Problem of MVDC SPS}
\label{sec:formulation}
In this paper, with the objective of maximizing the power delivering to loads, we formulate the problem as a MINLP subjected to operation constraints. The constrains are divided into AC, converter, DC zone, and load constraints described in the former subsections. Specifically, the objective function and constrains are expressed as
\begin{align}
\label{Objective} 
& \underset{ \{\mathcal{S}, \bm P_{g}, \bm Q_{g} \} } \max \; \sum_{l=1}^{L}w_{l}\cdot{s}_{l} P_{l} \\
& \text{s.t.} \quad (\ref{P_flow})-(\ref{PC_limit}), (\ref{eqn:redundence_switch}), (\ref{P_DC_cal}), (\ref{U_cons})-(\ref{eqn:vital_load}).
\end{align}
where $\bm P_{g} = \{P_{g_1}, \cdots, P_{g_M} \} $, and $\bm Q_{g} = \{Q_{g_1}, \cdots, Q_{g_M} \} $. $w_l \in \{ w_{V_1}, w_{V_2}, W_{V_3} \}$ and $s_l \in \mathcal{S} $ denotes the weight factor and switch status of the load $l$. The weight factors of three type loads are represented by $ w_{V_1}$, $w_{V_2}$, and $w_{V_3} $.


\section{System Management and Reconfiguration}
\label{sec:reconfiguration}
\begin{figure}[ht]
\begin{center} 
\setlength{\belowcaptionskip}{-0.3cm}
\includegraphics[width= 0.5 \textwidth]{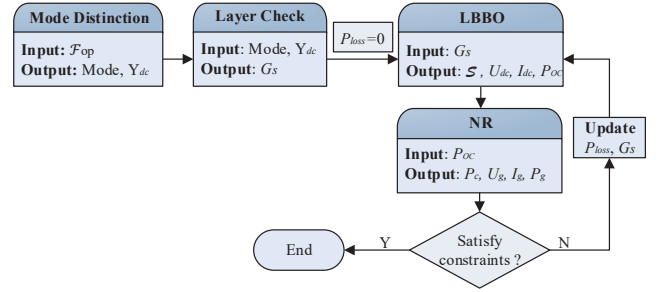} 
\caption{ NRBBO flowchart. } 
\captionsetup{justification=centering} 
\label{fig:flowchart}
\end{center} 
\end{figure}

Since the MVDC SPS is an AC/DC hybrid system, an NRBBO method is proposed to solve AC power flow and DC load allocatoin separately by updating $P_{loss_m}$. Considering the objective of restored load maximization, DC power flow is calculated by equation (\ref{total_dcpower}) before AC part. Firstly, fault mode distinction is executed according to the fault positions. Secondly, the layer number $ G_s $ where the search starts is checked. Thirdly, BBO algorithm is applied to calculate the parameters of DC part, such as switch set $ \mathcal{S} $ and voltage $ U_{dc_i} $. This algorithm is based on BBO because of its good performance in high dimensional optimization. Then, NR iteration is used to obtain the converter parameters. So the generators parameters can be obtained. At last, if the parameters of converters and generators don't satisfy the constraints, the $ P_{loss_m} $ and $ G_s $ will be updated to recalculate. The constraint (\ref{eqn:out_power}) in next BBO iteration will update in turn, thus optimal switch set $ \mathcal{S}_o $ can be gradully obtained. Fig.\ref{fig:flowchart} is the flow chart of the proposed method.


\subsection{Mode Distinction}
\begin{figure}[ht]
\begin{center} 
\setlength{\belowcaptionskip}{-0.3cm}
\includegraphics[width= 0.5 \textwidth]{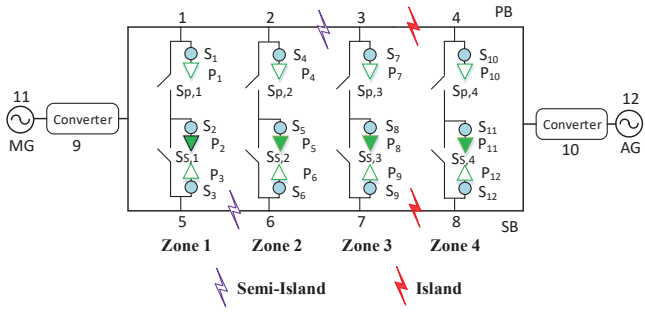} 
\caption{ Semi-Island and Island scenario. } 
\captionsetup{justification=centering} 
\label{fig:island}
\end{center}
\end{figure}

In order to accurately and effectively calculate DC power flow for reconfiguration, we divide all the fault scenarios into three modes, i.e., non-island fault, island fault and semi-island fault. The mode distinction algorithm is shown in Algorithm \ref{alg:mode_distinction}.
\begin{algorithm} [ht]
\caption{MODE Distinction}
\begin{algorithmic}
\IF {More than one fault in PB and SB respectively}
  \IF {Two faults in PB and SB of one zone}
    \STATE Mode $ = $ island;
  \ELSE
    \STATE Mode $ = $ semi-island;
  \ENDIF
\ELSE
  \STATE Mode $ = $ non-island;
\ENDIF
\end{algorithmic}
\label{alg:mode_distinction}
\end{algorithm} 
If only one side bus (PB or SB) has fault, these cases are defined as non-island mode. In island mode, there are faults happened in same zone, which causes the system is divided into two part. One case is shown in Fig. \ref{fig:island}. If there are more than two faults which happen in different zone, the system is also divided into two part. But there are coupled part between the two part. For example, in Semi-Island fault of Fig. \ref{fig:island}, zone 2 is the couple part, and the $S_{S_2}$, $S_{P_2}$ are the coupled redundancy switches. If $S_{S_2} = 0 $, $S_{P_2}=1$, the vital and semi-vital loads in zone 2 are powered by left island part connected to MG. Otherwise, powered by right island part connected to AG. 

In island mode as shown in Fig.\ref{fig:island}, the DC power grid has been divided into two independent parts. The redundancy switches do not need to be changed, and it is just need to modify the admittance $ \mathrm{Y}_{dc} $. If fault position is between bus 1 and 2, $Y_{dc_{12}}$ is set equal to zero. In each isolated part, the bus which connect to converter is selected to be slack bus. And the DC power flow of each part is calculated separately based on (\ref{total_dcpower}). 

In semi-island mode, the orignal structure is damaged. The coupled redundancy switches $S_{P_k}, S_{S_k}, k \in \Omega_c$ are need to be reconfigured. $\Omega_c \in \mathcal{K}$ denotes the set of coupled zone numbers. Thus the control vector contains $\mathcal{S}$, $\bm P_{g}$, $\bm Q_{g}$, and $S_{P_k}, S_{S_k}, k \in \Omega_c$. In order to reduce the computational complexity, two loop search is employed: the outer heuristic search is used to find optimal coupled redundancy switch reconfiguration due to the small scale variables; the inner search is used for optimal load shedding, which is described in the following subsection. Additionally, the corresponding admittences in $\rm{Y}_{dc}$ need to be modified according to the fault positions.

In non-island mode, the redundancy switches in the damaged sides are reconfigured to connect to another side bus. In order to avoid the over-voltage or over-current at the DC buses, the remaining redundancy switches $S_{P_k}, S_{S_k}, k \in \Omega_{ud}$ are need to be reconfigured. $\Omega_{ud} \in \mathcal{K}$ denotes the set of un-damaged zone numbers. The latter process of this mode is similar with the semi-island mode.

\subsection{Layer Search based on BBO Algorithm}
\label{subsec:layer_search}
When mode distinction is finished, layered search method is employed to calculate the DC power flow according to the priority of loads. Here the layer number $ G=3 $. The search space is reduced from $2^L$ to $G \cdot 2^ \frac{L}{3}$ if each priority loads has same quantity. Certainly, the layer number $ G $ can be determined by the requirement of actual system. The sets are searched according to the priority from high to low. If constraints check based on (\ref{total_dcpower}) are passed, skip ahead to search the next level set directly.

To distinguish the priority of loads, the weight factors must satisfy the following constraints:
\begin{align}
\label{Wsv} & w_{V_2} > \dfrac{w_{V_3} \cdot P_{V_3}}{P_{V_2}}, P_{V_3} \in \mathcal{P}_{NV}, P_{V_2} \in \mathcal{P}_{SV},\\
\label{Wv} & w_{V_1} > \dfrac{w_{V_2} \cdot P_{V_2}}{P_{V_1}}, P_{V_1} \in \mathcal{P}_{V},
\end{align}
where $ \mathcal{P}_{NV} $ , $ \mathcal{P}_{SV} $ and $ \mathcal{P}_{V} $ are the sets of non-vital, semi-vital, vital loads respectively. $ w_{NV} $ , $ w_{SV} $ and $ w_V $ denote the weight factors of three level loads respectively. 
Then, the lower bound of weight factors determined by (\ref{Wsv})-(\ref{Wv}) can be expressed as
\begin{align}
\label{Wsv_cal} & w_{V_2} > \dfrac{w_{V_3} \cdot P_{V_3}^\mathrm{max}}{P_{V_2}^\mathrm{min}}, \\
\label{Wv_cal} & w_{V_1} > \dfrac{w_{V_2} \cdot P_{V_2}^\mathrm{max}}{P_{V_1}^\mathrm{min}},
\end{align}
where $ w_{V_3} = 1 $ is considered as reference value, $ P_{V_3}^\mathrm{max} $ and $ P_{V_2}^\mathrm{max} $ represent the maximum element of $ \mathcal{P}_{NV} $ and $ \mathcal{P}_{SV} $ sets respectively, $ P_{V_2}^\mathrm{min} $ and $ P_{V_1}^\mathrm{min} $ the minimum element of $ \mathcal{P}_{SV} $ and $ \mathcal{P}_{V} $ sets respectively.

In order to solve the mixed integer programming (MILP) problem, BBO algorithm is employed in the layer search.
BBO algorithm is an evolutionary algorithm (EA) proposed by Dan Simon in 2008 \cite{simon2008biogeography}. The concept was motivated by biogeography based on migration and mutation of the distribution of biological species through time and space. 
As the major concept in BBO algorithm, migration and mutation are discussed below.

\subsubsection{Migration}
BBO algorithm is a population-based optimization algorithm where the population (habitat in BBO) is a set of candidate solutions. The goodness of candidate solutions are evaluated by the habitat suitability index ($ HSI $). Higher $ HSI $ means the solutions have better quality in the optimization problem. Features correlated with $ HSI $ include topographic diversity, land area and so on. Each of these features is called a suitability index variable ($ SIV $).
Emigration and immigration are used to probabilistically exchange information between solutions. Specifically, immigration rate $ \lambda $ is used to probabilistically determine whether modifying each $ SIV $ or not in a solution. Migration rate $ \mu $ of other solutions are used to probabilistically determine which one among the solution set will emigrate.
A solution with high $ HSI $ has abundance of species, so its emigration rate $ \mu $ is correspondingly large. Since the habitat has finite environment resources for further immigration, 
its immigration rate $ \lambda $ is small. 
$ \mu $ and $ \lambda $ of the $h$-th habitat can be expressed
as
\begin{align} 
\label{em_rate} &\mu_{h} = E\cdot\dfrac{h}{H},\\
\label{im_rate} &\lambda_{h} = A(1-\dfrac{h}{H}),
\end{align}
where $ E $ denotes the maximum emigration rate, $ A $ the maximum immigration rate, $ H $ the maximum habitat count.
In the case $ E=A $, the equations above result in
\begin{align}
&\mu_{h} + \lambda_{h} =E.
\end{align}

\subsubsection{Mutation}
Due to cataclysmic events, the $ HSI $ of a natural habitat may be changed drastically. In BBO when this event happens, the $ SIV $ mutates by the mutation rates. Hence the solutions have a chance to be better than their previous values. The mutation scheme tends to increase diversity among the solutions. But mutation operation is a high risk process, solutions probabilistically become inferior after mutation process than the previous. Mutation scheme has many kinds of implementations, such as replacing with randomly generated solution, mutation like GA and so on.

In migration and mutation process, a few elite solutions are kept in BBO to prevent the best solutions from being damaged.

\begin{figure}[ht]
\begin{center} 
\setlength{\belowcaptionskip}{-0.3cm}
\includegraphics[width=0.4 \textwidth]{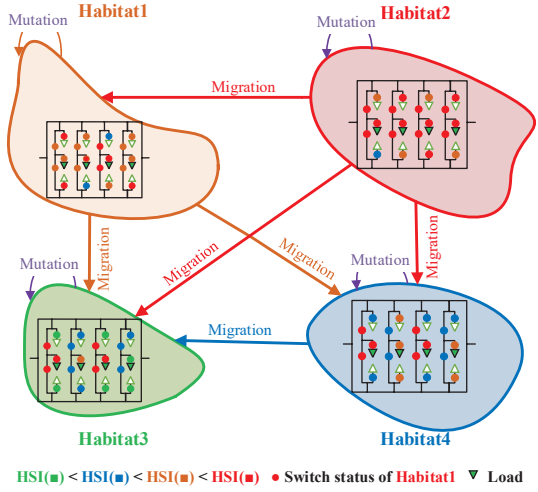} 
\caption{ BBO algorithm for reconfiguration. } 
\captionsetup{justification=centering} 
\label{fig:bbo_in_sps}
\end{center} 
\end{figure}
BBO algorithm is used to solve DC zone power flow problem and obtain optimal solutions satisfying zonal constrains. In this section it is introduced in detail. Fig.\ref{fig:bbo_in_sps} shows the migration and mutation processes of BBO for reconfiguration.
In the formulated problem, $ SIV $ represents the switch status $ s_{l} $, so the $ h $-th habitats $ \mathcal{H}_{h} $ can be expressed as
\begin{equation}
\begin{aligned}
\mathcal{H}_{h} = & [ SIV_{h,1}, SIV_{h,2}, \cdots, SIV_{h,Z} ]
= & [S_{h,1}, S_{h,2}, \cdots , S_{h,L}],
\end{aligned}
\end{equation}
where $ SIV_{h,z} $ represents the value of the $ z $-th independent variable in the $ h $-th habitat. 

In the previous section, the power equation in DC zone is written as (\ref{P_DC_cal}). Due to that zone power is calculated in the external iteration firstly, $ P_{oc_m} $ is needed. However, $ P_{loss_m} $ calculated by (\ref{P_loss}) is still unknown in the initial state, we set it to zero. The lines between generators and converters are also assumed to be lossless. So the initial $ \sum_{m =1}^{M} P_{oc_m} $ equals to $ \sum_{m=1}^{M} P_{G_m} $.
The detail process of BBO for reconfiguration in DC zone is introduced as follows. 

First, the solutions ${H}_{h}$ are generated by the search start layer $G_s$. If $G_s = 3$, the random solutions only generate in switches of non-vital loads while vital and semi-vital loads are all set to work.

Second, the feasibility of generated solutions are checked by (\ref{U_cons})-(\ref{I_cons}). In each solution the load power of each bus can be calculated by the load switch configuration. The voltage and current can be checked after calculation of dc power flow based on (\ref{P_DC_cal}). 
If a solution does not pass the feasibility check, mutation operation is carried on until a feasible one is generated, which is different from the traditional BBO algorithm. The migration operation is carried out after feasibility check.  

Third, we check whether the algorithm would converge when finish the one generation calculation. 
Due to weight factor, the cut-off conditions of layered search cannot be the same. $ O_g $ denotes the cut-off error between the best solutions of two generations. According to (\ref{Wsv_cal})-(\ref{Wv_cal}), it can be defined as
\begin{equation}
\begin{aligned}
O_{g} = w_{V_g} P_{V_g}^ \mathrm{min}, g \in \{1,2,3\},
\end{aligned}
\end{equation}
where $ w_{V_g} \in \{w_{V_1}, w_{V_2}, w_{V_3}\}$, $ P_{V_g}^\mathrm{min} \in \{ P_{V_1}^\mathrm{min}, P_{V_2}^\mathrm{min}, P_{V_3}^\mathrm{min}\}  $.
In order to keep enough search, if the objective error $ e $ of the best solutions among two generations is lower than $ O_g $, we also execute $ RE $ iterations.

In \cite{ma2014convergence}, the result indicates that when BBO algorithm is used in a binary search problem, the elite version that maintains the best solution can converge to a solution subset, which contains one global optimal solution. Our layered search and mode distinction will reduce the search space of switch variables and increase the mutation probability. It does not change the convergence characteristic of BBO.

\subsection{Converter Power Loss and AC Power Flow}

\begin{algorithm}
	\caption{The proposed hybrid method-NRBBO}
	\begin{algorithmic}[1]
		\REQUIRE ~\\
		The fault bus number set: $ \mathcal{F}_{po} $;
		\ENSURE~\\
		$ \mathcal{S} $, $ P_{g_m}$, and $Q_{g_m} $.\\
		\STATE MODE distinction by Algorithm \ref{alg:mode_distinction}; \\
		\STATE Layer check by constraints (\ref{eqn:redundence_switch}), (\ref{P_DC_cal});\\
		\STATE Generate $ H $ habitats by $ G_s $;
		$ P_{loss_m} \leftarrow 0 $; \\
		\REPEAT
		\FOR { $ g \leftarrow G_s  $ to $ G $}
		\FOR { $ h \leftarrow 1 $ to $ H $}
		\STATE Calculate the objective value of solutions, keep 2 elite solutions, and operate migration by $ \lambda_h $ and $ \mu_h $;
		\STATE Bus status calculation by MODE and (\ref{P_DC_cal});
		\WHILE {Not satisfy constraints (\ref{U_cons}) - (\ref{I_cons}) }
		\STATE Generate a new habitat that satisfy (\ref{U_cons}) - (\ref{I_cons});
		\ENDWHILE
		\ENDFOR
		\IF {$ e < O_g $ and $ r \geq RE $}
		\STATE break;
		\ENDIF	
		\ENDFOR \\
		Calculate $ P_{C_m} $, $ P_{loss_m} $, $ P_{g_m} $, and $ Q_{g_m} $ by (\ref{NRLoop}), (\ref{U_gen})-(\ref{S_line});
		\STATE Update $ P_{loss_m} $, $ G_s \leftarrow g $;
		\UNTIL{Satisfy (\ref{PG_limit})-(\ref{delta_limit}) and (\ref{IC_limit})-(\ref{PC_limit})}
	\end{algorithmic}
	\label{alg:total_alg}
\end{algorithm}

In this part, at first the parameters of converter are calculated by the $P_{oc_m}$ obtained in subsection \ref{subsec:layer_search}, then the generator's can be determined by the voltage drop and power loss equations of transmission line while keeping it in the restricted range. 
ßß
According to the equations (\ref{P_loss}), the converter loss depend on the input current magnitude $ I_{C_m} $. Because DC network is calculated in the first step, $ I_{C_m} $ is unknown. In this part, $ P_{oc_m} $ obtained in the former part are constant parameters. 
In order to calculate $P_{loss_m}$ , an Newton-Raphson iteration based on $P_{C_m}$ and $Q_{C_m}$ is used. $ V_{C_m} $ and $ \theta_{C_m} $ are kept constant.
 For each converter, the iteration is updated as follows
\begin{equation}
\begin{aligned}
\label{NRLoop} {f_m}^{(t)} = & - \left[ \left(\dfrac{\partial {f}}{\partial P_{C_m}}\right)^{(t)} \: \left(\dfrac{\partial {f}}{\partial Q_{C_m}}\right)^{(t)} \right]  \\
& \cdot \left[ {\Delta P_{C_m}}^{(t)} \: {\Delta Q_{C_m}}^{(t)} \right]^ \mathrm{T},
 \end{aligned}
\end{equation}
with the function $f_m$ given by
\begin{equation}
\begin{aligned}
{f_m}^{(t)} & = {P_{C_m}}^{(t)} - P_{oc_m} - {P_{loss_m}}^{(t)},
\end{aligned}
\end{equation}
where $ t $ is the iteration index of Newton-Raphson method.
After the convergence of $ f_m $, the active and reactive power of generators can be calculated by (\ref{U_gen})-(\ref{S_line}). 
\begin{equation}
\begin{aligned}
\label{U_gen} {U}_{G_m}\angle{\delta_G}_m & = {U}_{C_m}\angle{\delta_C}_m + \dfrac{P_{C_m} R_m+Q_{C_m}X_m }{U_{C_m}}  \\
& + j\dfrac{P_{C_m} R_m+Q_{C_m}X_m }{U_{C_m}},
 \end{aligned}
\end{equation}

\begin{align}
\label{S_line} & \Delta P_{ln_m} + \Delta Q_{ln_m} =  \dfrac{P_{C_m}^{2} + Q_{C_m}^{2}}{U_{C_m}^{2}}(R_m+jX_m),
\end{align}
where $ {U}_{G_m} $, $ {\delta_G}_m $ denote the voltage and angle of generator, $ {U}_{C_m} $, $ {\delta_C}_m $ the voltage and angle of converter, $ \Delta P_{ln_m} $, $ \Delta Q_{ln_m} $ the active and reactive power loss in transmission line.

At last, AC constraints (\ref{PG_limit})-(\ref{delta_limit}) are checked in the overall iteration loop. 
The detail NRBBO algorithm is shown in Algorithm \ref{alg:total_alg}.


\section{Simulations}
\label{sec:simulation}
A MVDC SPS with $ K = 6 $, $ N=14 $ and $ M = 2 $ is shown in Fig. \ref{fig:normal}. Six DC load zones are fed power from one MG and AG. This model is used for validation and analysis of our algorithm. The simulation parameters of power network are chosen by shipboard power requirements of IEEE Std 1709 \cite{ieee2010mvdc}. 
The detail parameters in MVDC SPS model are shown in Table \ref{tab:Simulation_parameters}. The algorithm parameters are set as follows. The weight factors are set as $ w_{V_1} = 12 $, $ w_{V_2} =4 $, $ w_{V_3}=1 $ by (\ref{Wsv_cal})-(\ref{Wv_cal}). Here $ E $ and $ I = 1 $ and keep elitism $ = 2 $. 

\begin{table}[!htbp] \centering
\caption{Simulation Parameters}
\label{tab:Simulation_parameters}
\begin{tabular}{|c|c|c|}
\hline
Parameters & Max. & Min. \\
\hline
$ PG_{i}(MG) $ & 8MW & 0MW \\
\hline
$ PG_{i}(AG) $ & 4MW & 0MW \\
\hline
$ V_{ac} $ & 3.49kV & 2.97kV  \\
\hline
$ \delta_{i} $ & 1 & -1  \\
\hline
$ V_{dc} $ & 1.1kV & 0.9kV  \\
\hline
\end{tabular}
\end{table} 


\begin{table}[!htbp] \centering 
\label{tab:load_parameters}
\setlength{\belowcaptionskip}{-0.3cm}
\caption{Loads in 6 Zone SPS}
\begin{tabular}{|c|c|c|c|c|c|c|}
\hline
\multirow{2}{*}{Power \& Number} & \multicolumn{6}{c|}{Zone No.} \\ \cline{2-7}
              & 1 & 2 & 3 & 4 & 5 & 6 \\ \hline
$ P_{VL} $(MW) & 0.2$ \times2 $ & 0.5$ \times2 $ & 0.3$ \times2 $ & 0.5$ \times2 $ & 0.8$ \times2 $ & 0.3$ \times2 $ \\
\hline
$ P_{SVL} $(MW) & 0.4$ \times2 $ & 0.3$ \times2 $ & 0.3$ \times2 $ & 0.2$ \times2 $ & 0.2$ \times2 $ & 0.4$ \times2 $ \\
\hline
$ P_{NVL} $(MW) & 0.2$ \times 2 $ & 0.1$ \times 2 $ & 0.2$ \times 2 $ & 0.2$ \times 2 $ & 0.1$ \times 2 $ & 0.2$ \times 2 $ \\
\hline
\end{tabular}
\end{table}

\subsection{Performance of NRBBO }
\begin{figure}[!htbp]
\begin{minipage}[t]{0.5\linewidth}
\centering
\includegraphics[width=1 \textwidth]{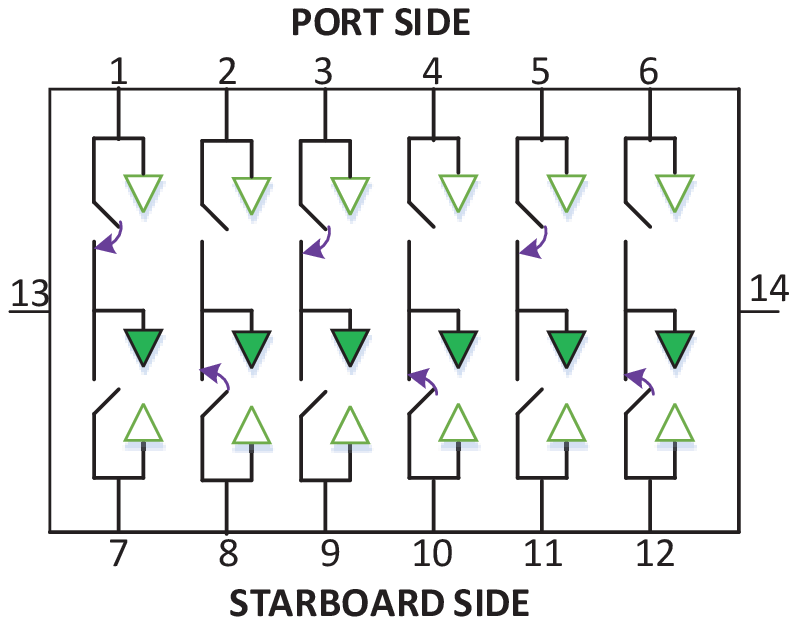}
\centerline{(a)}
\label{fig:scenario:a}
\end{minipage}%
\begin{minipage}[t]{0.5\linewidth}
\centering
\includegraphics[width=1 \textwidth]{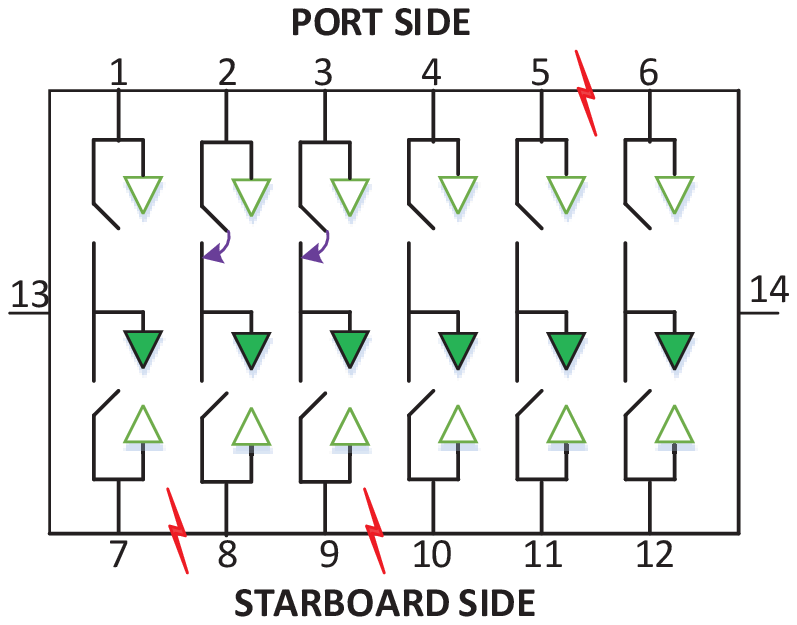}
\centerline{(b)}
\label{fig:scenario:b}
\end{minipage}
\caption{Initial and fault scenario.}
\label{fig:scenario:one}
\end{figure}

The system work in an optimal configuration under normal condition, in which all the loads are powered to their full capacities. Fig. \ref{fig:scenario:one}(a) shows a pre-fault condition where all the loads are serviced for a particular switch configuration. After a fault happens, the power can be restored by result of reconfiguration method in a optimal or suboptimal status.
Now a fault scenario is considered where faults occur as shown in Fig. \ref{fig:scenario:one}(b). The portion of the PB between $7$ and $10$ is thus left without power and the configuration of switches needs to be changed so that the loads can be serviced based on their priorities. 
The restored power is shown in Table \ref{tab:load_restored}.
It can be observed that best solution is found by our algorithm. Due to the stochastic characteristic of our algorithm in DC part, near-optimal solutions are obtained in the most time. The power of MG and AG drops to $5.85$ MW and $3.78$ MW in the best solution. 
Table \ref{tab:switch_change} also shows the switch status of best solution.
In this Table, S2(5/11) denote the switch of semi-vital load $2$ between bus $5$ and $11$, N1,2(1,2) the four switches of non-vital loads $1$ and $2$, which connect to bus $1$ and $2$.


\begin{table}[ht]
\centering
\caption{Load Restored}
\label{tab:load_restored}
\begin{tabular}{|c|l|l|l|l|l|l|l|}
\hline
\multicolumn{2}{|l|}{}        & \multicolumn{1}{c|}{$ P_{total} $} & \multicolumn{1}{c|}{$ P_{loss} $} & $ P_{G_1} $ & $ P_{G_2} $ & $ P_{C_1} $ & $ P_{C_2} $ \\ \hline
\multicolumn{2}{|l|}{Initial} & 11.6 & 0.34 & 7.94 & 3.998 & 7.91 & 3.99 \\ \hline
\multirow{3}{*}{Final}  & Best& 9.5  & 0.30 & 5.98 & 3.82 & 5.95  & 3.81  \\ \cline{2-8} 
                   & Mean     & 9.33 & 0.29 & 5.85 & 3.78 & 5.83  & 3.77 \\ \cline{2-8} 
                   & Worst    & 9    & 0.29 & 5.57 & 3.72 & 5.55  & 3.71 \\ \hline
\multicolumn{2}{|l|}{LINGO}   & 9.5  & 0.30 & 5.98 & 3.82 & 5.95  & 3.81 \\ \hline
\end{tabular}
\end{table}

\begin{table}[ht]
\centering
\caption{Switch Change}
\label{tab:switch_change}
\begin{tabular}{|c|l|l|l|l|l|l|l|}
\hline
\multicolumn{2}{|c|}{}  & \multicolumn{3}{c|}{Zone} & \multicolumn{3}{c|}{Load} \\ \hline
\multicolumn{2}{|c|}{Switch No.}  & 1 & 2 & 6 & S2(5/11) & N1,2(1,2,3,4,5,7,10) & N2(11)    \\ \hline
\multirow{2}{*}{Status} & Initial & 1 & 0 & 0 & 1 & 1 & 1   \\
\cline{2-8} 	        & Best    & 0 & 1 & 1 & 0 & 0 & 0     \\ \hline
\end{tabular}
\end{table}


\begin{figure}[!htbp]
	\setlength{\belowcaptionskip}{-0.3cm}
	\begin{minipage}[t]{0.5\linewidth}
		\centering
		\includegraphics[width= 1 \textwidth]{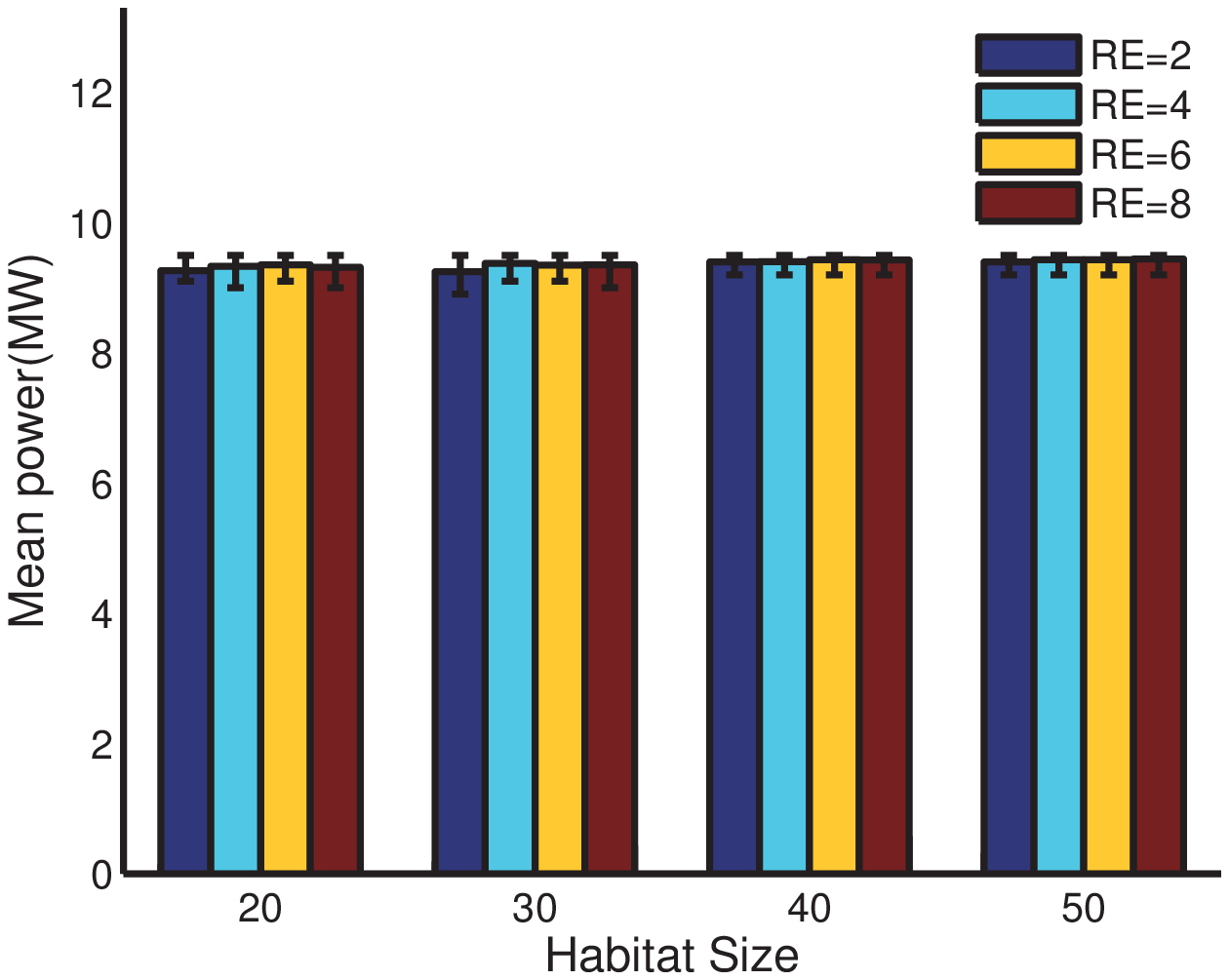}
		\centerline{(a)}
		\label{fig:performance:a}
	\end{minipage}%
	\begin{minipage}[t]{0.5\linewidth}
		\centering
		\includegraphics[width= 1 \textwidth]{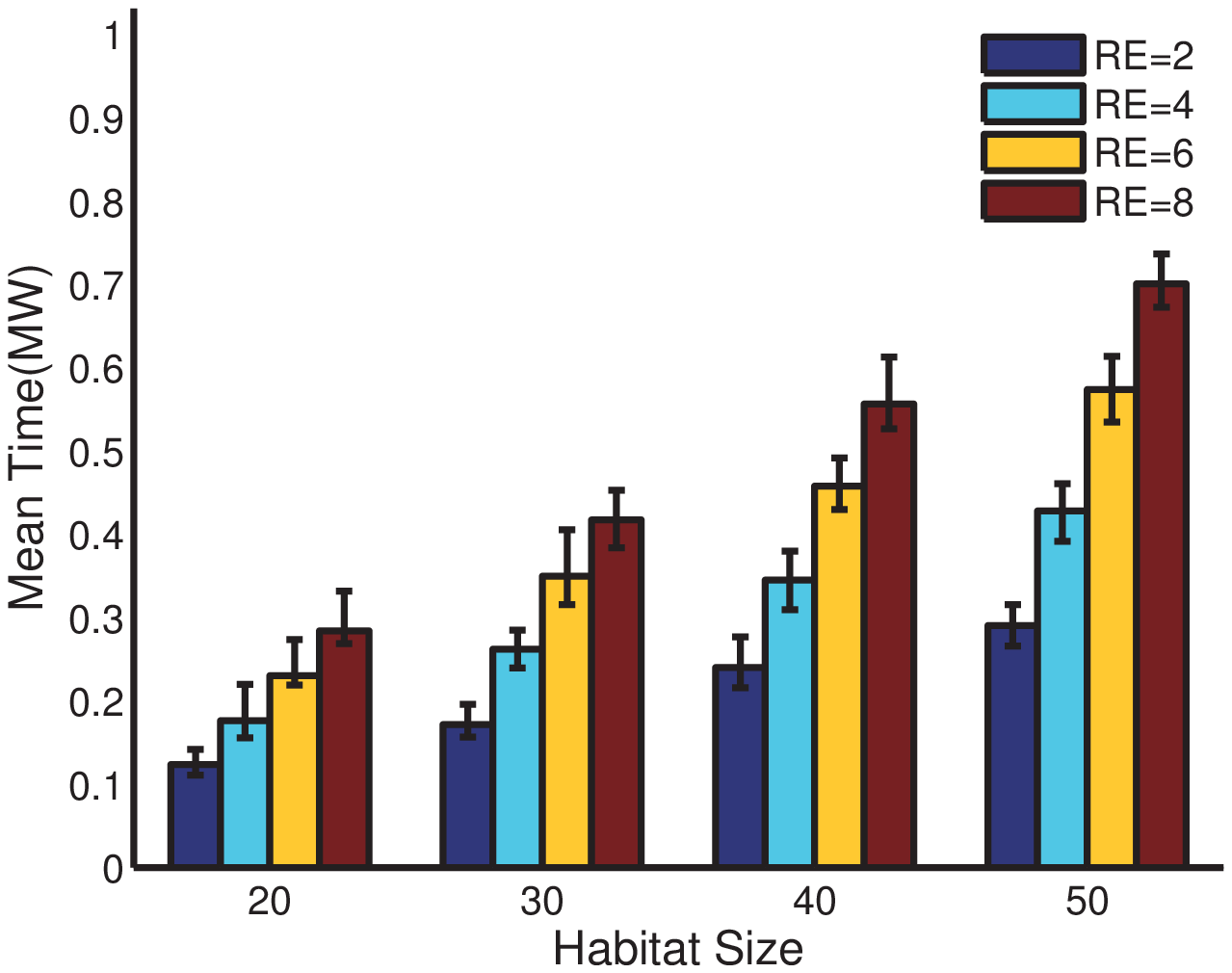}
		\centerline{(b)}
		\label{fig:performance:b}
	\end{minipage}
	\caption{ (a) Average restored load power and error.  (b) average run time and error.}
	\label{fig:performance:one}
\end{figure}

%

Fig. \ref{fig:performance:one} present results of sensitivity analysis of NRBBO with different parameters $ H $ and $ RE $. These include best, worst and mean value. Different with most evolutionary algorithm, $ H $ and $ RE $ have a little effect on the performance of restored power in our algorithm.
It is clear that the increasing the size of solution set will increase the exploration during simulation but at the cost of execution time. Moreover, increasing the number of iterations will not necessarily improve the performance as the objective value will converge after certain iterations.

\subsection{Comparison With Other Evolutionary Algorithms}

\begin{figure}[!htbp]
	\setlength{\belowcaptionskip}{-0.3cm}
	\begin{minipage}[t]{0.5\linewidth}
		\centering
		\includegraphics[width= 0.99\textwidth]{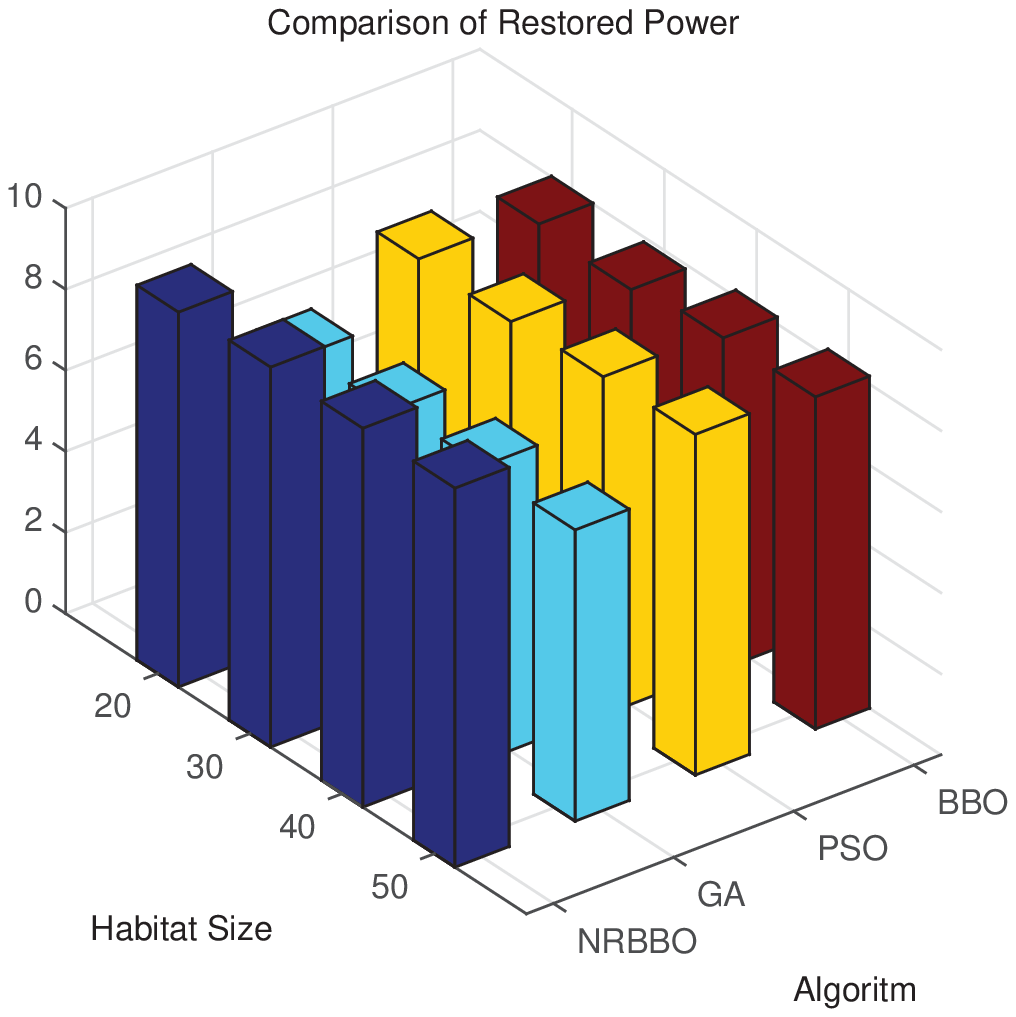}
		\centerline{(a)}
		\label{fig:comparison:a}
	\end{minipage}%
	\begin{minipage}[t]{0.5\linewidth}
		\centering
		\includegraphics[width=0.99\textwidth]{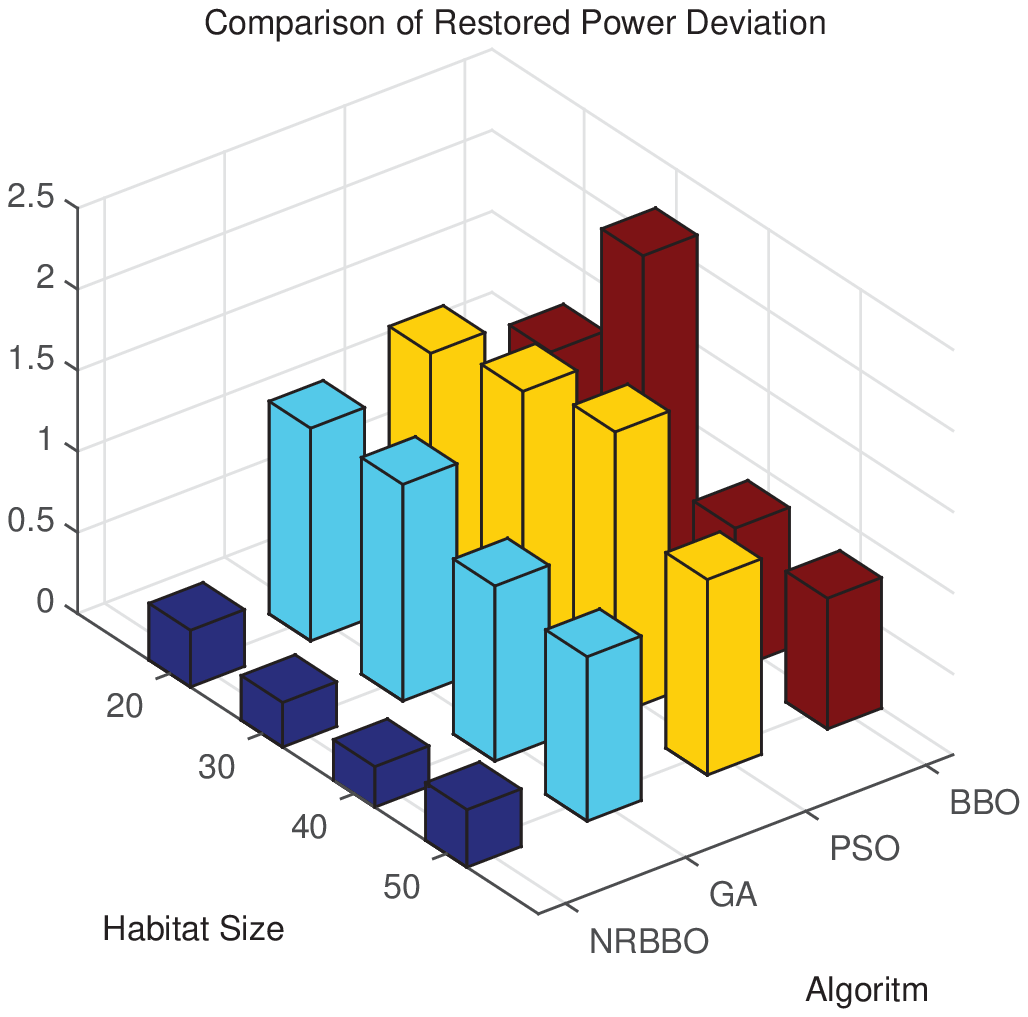}
		\centerline{(b)}
		\label{fig:comparison:b}
	\end{minipage}
	\caption{ (a) Average restored power comparison.  (b)  average restored power deviation comparison.}
	\label{fig:comparison_one}
\end{figure}

%
%

Since the goal of this study is to design an algorithm to realize fast reconfiguration in MVDC SPS, a comparison of the performance of NRBBO between the traditional BBO, PSO and GA based algorithms is given including restored power and execution time.
The switches status within our proposed algorithm depends on stochastically generated variable. Therefore, to 
improve the reliability of the conclusions about performance, simulations for 50 consecutive runs were carried out for all the four algorithm. The habitat size (population size)
and the maximum generation $ N_g $ were kept same for all the algorithms to make better comparison of the results.

Fig. \ref{fig:comparison_one}  show the comparison of the average restored power and its deviation among NRBBO, BBO, PSO and GA based algorithms with different repetition $ RE $ and habitat size $ H $. The comparisons demonstrate that our algorithm can find a better solution, which can restored more loads than others. The deviation of solutions found with NRBBO is also lower than the others. Lower deviation value of restored power suggests our algorithm performs better stability. 

\begin{figure}[!htbp]
	\setlength{\belowcaptionskip}{-0.3cm}
	\begin{minipage}[t]{0.5\linewidth}
		\centering
		\includegraphics[width=1.1 \textwidth]{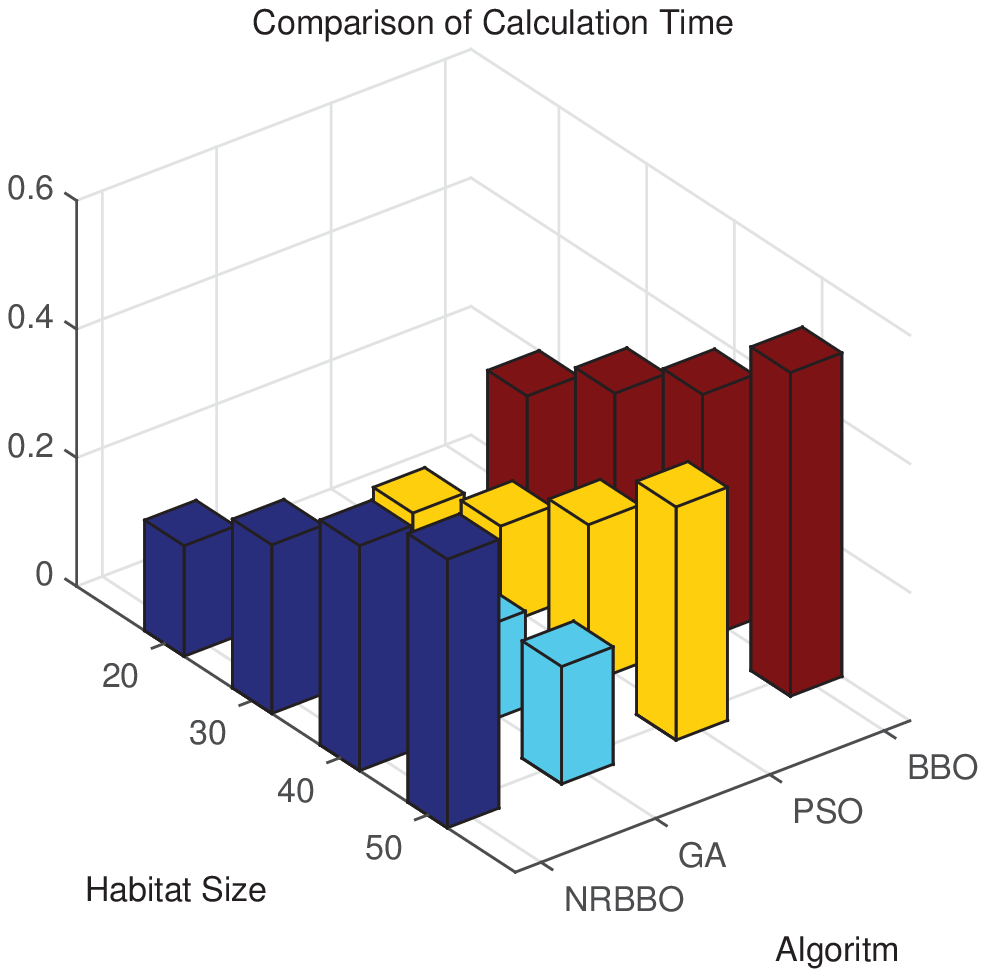}
		\centerline{(a)}
		\label{fig:comparison_two:a}
	\end{minipage}%
	\begin{minipage}[t]{0.5\linewidth}
		\centering
		\includegraphics[width=1.1 \textwidth]{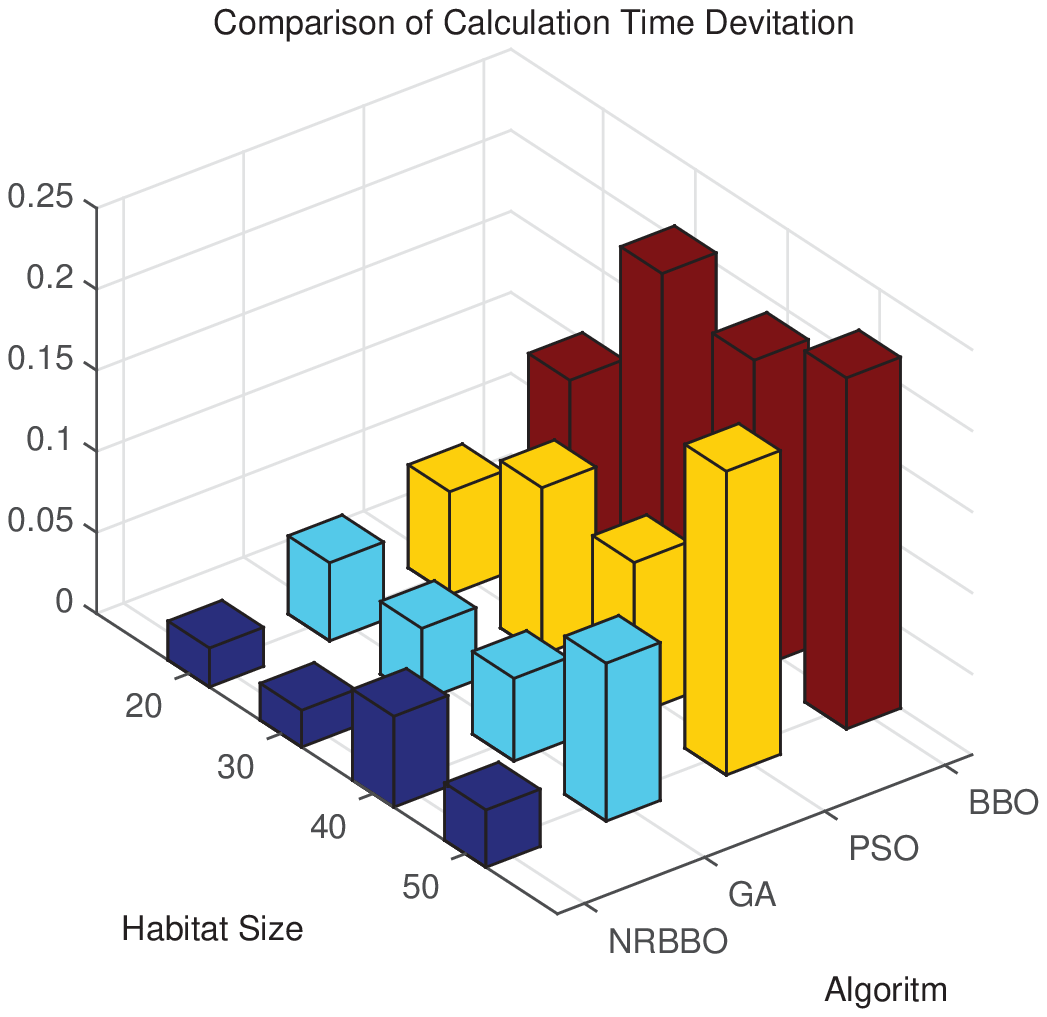}
		\centerline{(b)}
		\label{fig:comparison_two:b}
	\end{minipage}
	\caption{ (a) Average calculation time comparison. (b)  calculation time deviation comparison.}
	\label{fig:comparison_two}
\end{figure}

%
The comparison of calculate time is shown in Fig. \ref{fig:comparison_two}. 
We can know that the execution time of our algorithm is less than traditional BBO and PSO based algorithm, and it's close to GA based algorithm. But our stability of execution time is better than the others. In summary, the algorithm that we proposed has good performance in restore power and execution time. At the same time, it performs better stability than the others in this problem.




\subsection{Comparison With Other Reconfiguration Methodologies}
The execution time of an algorithm is determined by its computational complexity. In this subsection, the complexity of the methods provided to realize optimal reconfiguration of the MVDC SPS is analyzed. The former classic methods include branch-and-bound method ("LINGO" software), Interior-point method by combining Newton's method\cite{Bose2012Analysis}, Reinforce learning\cite{das2013dynamic}.
The Interior-point method solves the reconfiguration problem by applying Newton's method to a sequence of equality constrained problems. The worst-case complexity for interior-point based method is more than $ O(Ld^2) $, where $ L $ is the number of loads, $ d $ is the number of constraints. 
Reinforcement learning based reconfiguration method uses greedy strategy to exploration. The learning process will increase the complexity. So the complexity is greater than $ O(2^L) $. 
 
\begin{table}[ht]
\centering
\caption{Comparison of Algorithm Complexity}
\label{tab:algorithm_comparison}
\begin{tabular}{|l|l|}
\hline Algorithm & Complexity \\
\hline LINGO  & $ O(2^L)  $ \\
\hline Das, Sanjoy \& Bose, Sayak \cite{das2013dynamic} & $ O(2^L) $ \\
\hline Bose, S \& Pal, S \cite{Bose2012Analysis} & $ O(Ld^2) $  \\
\hline NRBBO &  $ O( 2^{(\frac{L}{3})}) $ \\
\hline
\end{tabular} 
\end{table}
For NRBBO, the feasibility check of solutions to satisfy the constraints is the major contributor to computational complexity. The execution time is directly proportional to the number of the feasibility check, which is correlation to the habitat size. Due to the layered search and mode distinction methods used in our algorithm, the maximum search space of switch variables is reduced to $ 2^{(\frac{L}{3})} $. The complexity comparison with the other methodologies in reconfiguration of MVDC SPS is shown in Table. \ref{tab:algorithm_comparison}.

\subsection{Relationship between Fault Position and Restored Power}
\begin{figure}[!htbp]
\setlength{\belowcaptionskip}{-0.3cm}
\begin{center} 
\includegraphics[height = 0.28 \paperwidth]{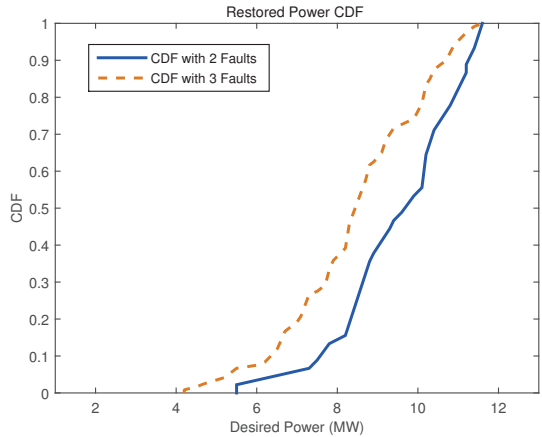} 
\caption{ Restore Power CDF.  } 
\captionsetup{justification=centering}
\label{fig:probability_analysis} 
\end{center}
\end{figure}

In this subsection, an analysis of the relationship between the fault position, number and restored power is illustrated by cumulative distribution function (CDF). It is defined as $ f(P_d) = Prob. \{P_r \leq P_d\} $, where $ P_d $ is the desired power. 
The CDF of the restored power to the loads is plotted for all the possible locations for two and three faults, which is shown in Fig. \ref{fig:probability_analysis}.

The restored power also can drop to less than 6 MW in the 2-fault and 3-fault scenarios, which can be observed from the CDF analysis of the system. In 2-fault scenarios, the vital loads can be always serviced. But in 3-fault scenarios, there are less than 10 \% probability that the vital loads cannot be fully seriviced. The restored power is not only correlation with numbers of fault, but also related to the position. The line faults impact on the restored power are greater when they are closer to the MG. For example, if the faults happen in 1-2 and 7-8 buses, the MG just need to deliver power to one zone, and the power will drop greatly. Therefore, the transmission lines that close to MGs need to be more protected.




\section{Conclusion}
\label{sec:conclusion}
In this paper, comprehensive multi-zone MVDC SPS is modeled. The multi-zone DC power flow calculation and converter power model for reconfiguration is given. Additionally an hybrid optimization method NRBBO is proposed to solve reconfiguration problem with multi-constraints when the power network has faults in SPS. The performance of our algorithm is better than traditional evolutionary based methods for this problem. The results clearly demonstrate the stability that the solutions found by this method have small deviation. Besides, our algorithm has relative low complexity than other reconfiguration methods used in MVDC SPS. 
The future works include the optimization of minimizing power loss for energy saving, while maximizing the weighted loads, and application of this method in other SPSs.

\bibliographystyle{ieeetr}
\bibliography{Privacy}

\end{document}